\documentclass[twocolumn,aps,showpacs,eqsecnum]{revtex4}
\usepackage{amsmath}
\usepackage{amssymb}
\begin{document}
\title{Comparison of Quantum and Classical Local-field Effects on
Two-Level Atoms in a Dielectric}
\author{Michael E. Crenshaw}
\affiliation{AMSRD-AMR-WS-ST, US Army RDECOM, Aviation and Missile RDEC,
Redstone Arsenal, AL 35898, USA}
%\centerline{cygwin/home/max/manuscripts/ce19.tex}
\date{\today}% any date may be explicitly specified
\begin{abstract}
The macroscopic quantum theory of the electromagnetic field in a
dielectric medium interacting with a dense collection of embedded
two-level atoms fails to reproduce a result that is obtained from
an application of the classical Lorentz local-field condition.
Specifically, macroscopic quantum electrodynamics predicts that the
Lorentz redshift of the resonance frequency of the atoms will be
enhanced by a factor of the refractive index $n$ of the host medium.
However, an enhancement factor of $(n^2+2)/3$ is derived using the
Bloembergen procedure in which the classical Lorentz local-field
condition is applied to the optical Bloch equations.
Both derivations are short and uncomplicated and are based on
well-established physical theories, yet lead to contradictory results.
Microscopic quantum electrodynamics confirms the classical
local-field-based results.
Then the application of macroscopic quantum electrodynamic theory to
embedded atoms is proved false by a specific example in which both the
correspondence principle and microscopic theory of quantum
electrodynamics are violated.
\end{abstract}
\maketitle
\par
\section{Introduction}
\par
The effect of a dielectric host medium on the spontaneous emission rate
of a two-level atom remains an interesting and challenging problem in
quantum optics with an importance that befits a rigorous test of our
understanding of the interaction of light with matter.
The essential characteristic of the problem is its multi-scale nature
with the atom being a creature of microscopic quantum electrodynamics
and the dielectric manifesting in the realm of classical continuum
electrodynamics.
In principle, it is possible to represent both the atom and dielectric
microscopically although, in practice, the spontaneous emission rate is
calculated using a macroscopic quantum electrodynamic theory
\cite{BIGinz,BIJauchW,BIMarcuse,BIGlaub,BIMilJMO,BINien} in which the
continuous dielectric medium is incorporated into a medium-assisted
electromagnetic field.
While the dielectric renormalization of the spontaneous emission rate of
an embedded atom generates considerable interest, a mere
handful of papers discuss the effect of a dielectric on the Lorentz
redshift of the resonance frequency of two-level atoms.
Knoester and Mukamel \cite{BIKnoesM} used a Hopfield \cite{BIHop}
polariton model of macroscopic quantum electrodynamics and found that
the Lorentz redshift is enhanced by a factor of the refractive index $n$
compared to the vacuum value.
Crenshaw and Bowden \cite{BICBPRA96} derived the enhancement factor
$(n^2+2)/3$ using the Bloembergen \cite{BIBloem} procedure in which the
classical Lorentz local-field condition is applied to the 
optical Bloch equations.
The two procedures also produce different values for the renormalization
of the field (Rabi frequency) that drives the dynamics of atoms in a
dielectric.
\par
In this paper, we derive the generalized optical Bloch equations for a
dense collection of two-level atoms in a dielectric host material using,
first, the Bloembergen procedure based on the classical Lorentz
local-field condition and, second, Ginzburg macroscopic quantum
electrodynamics.
The derivations are simple and direct and based on well-established
physical theories, yet produce contradictory results for
(i) the Lorentz redshift of the resonance frequency of two-level
atoms densely embedded in the dielectric and
(ii) the renormalization of the field that drives the dynamics
of the embedded atoms.
The microscopic quantum electrodynamic procedure described by
Crenshaw and Bowden \cite{BICBPRL} confirms the
Bloembergen-based results.
Then, we must conclude that the macroscopic quantum electrodynamic
theory applied to dielectrically embedded atoms violates both the
correspondence principle and microscopic quantum electrodynamics.
\par
\section{Dielectric Local-field effects}
\par
In the view of Lorentz, classical continuum electrodynamics is better
expressed in terms of an atomistic model of discrete particles embedded
in the vacuum that interact with the microscopic electromagnetic field
at the point of the particle \cite{BIJackson,BIBorn}.
The local field that acts on that particle,
\begin{equation}
{\bf E}_L={\bf E}+\frac{4\pi}{3}{\bf P},
\label{EQq2.01}
\end{equation}
is comprised of the macroscopic Maxwell field ${\bf E}$ and the
reaction field of all other particles, expressed in terms of the
macroscopic polarization ${\bf P}$.
For a linearly polarizable material, the polarization
${\bf P}= p N_d {\bf E}_L$ is the
product of the microscopic polarizability $p$,
the dipole number density $N_d$, and the local field.
Using the Lorentz local-field condition (\ref{EQq2.01}) to eliminate the
microscopic local field produces the Clausius--Mossotti--Lorentz--Lorenz
relation
\begin{equation}
\frac{4\pi}{3}  pN_d =
\frac{\varepsilon-1}{\varepsilon+2}
\label{EQq2.02}
\end{equation}
between the polarizability and the macroscopic dielectric constant
$\varepsilon=1+4\pi P/E$.
\par
The local-field principle also applies to nonlinear media.
Bloembergen \cite{BIBloem} investigated nonlinear optics in the presence
of a linear host medium and found that
a local-field dielectric enhancement factor of
\begin{equation}
\ell=\frac{n^2+2}{3}
\label{EQq2.03}
\end{equation}
accompanies each appearance of a macroscopic field in the nonlinear
susceptibility.
In our notation, $\varepsilon=n^2$ for a dielectric and $\ell$ refers to
the specific quantity in the preceding equation and not to any other
constant or variable representing a local-field factor.
Bowden and co-workers \cite{BIBBE,BIIOB1,BIIOB2} predicted intrinsic
optical bistability in a dense collection of vacuum-embedded two-level
atoms due to an inversion-dependent local-field shift of the resonance
frequency.
Later work \cite{BICBPRA96} reported the effect of embedding the dense
two-level systems in a host dielectric.
The dynamics of the two-level systems are described by the
optical Bloch equations \cite{BIBoyd}.
Building on Bloembergen's work \cite{BIBloem}, Bowden and Dowling
\cite{BIBowDowl} showed that the field that drives the atoms is the
local field (\ref{EQq2.01}).
With that substitution, one obtains the generalized optical Bloch
equations \cite{BIBowDowl},
\begin{subequations}
\label{EQq2.04}
\begin{equation}
\frac{\partial R_{21}}{\partial t}
=i (\omega-\omega_0) R_{21} -
\frac{i \mu }{2\hbar}
\left ({\cal E}+{\frac{4\pi}{3}}{\cal P}\right ) W-\gamma_{\perp} R_{21},
\label{EQq2.04a}
\end{equation}
$$
{\frac{\partial W}{\partial t}}=
- {\frac{i\mu}{\hbar}} \left [ 
\left ({\cal E}^*+{\frac{4\pi}{3}}{\cal P}^*\right ) R_{21}
-\left ({\cal E}+{\frac{4\pi}{3}}{\cal P}\right ) R_{21}^* \right ] 
$$
\begin{equation}
-\gamma_{\parallel} (W-W_{\rm eq}) .
\label{EQq2.04b}
\end{equation}
\end{subequations}
Here, fields are represented in the plane-wave limit by envelope
functions such that
$P={\frac{1}{2}}({\cal P} e^{-i\omega t}+{\rm c.c.})$,
$E ={\frac{1}{2}}({\cal E} e^{-i\omega t}+{\rm c.c.})$,
and
$E_L ={\frac{1}{2}}({\cal E}_L e^{-i\omega t}+{\rm c.c.})$.
The macroscopic spatially averaged atomic variables in a rotating
frame of reference are
$R_{21}=\langle \rho_{21} e^{i\omega t} \rangle_{\rm sp}$,
$R_{12}=\langle \rho_{12} e^{-i\omega t} \rangle_{\rm sp}$,
and
$W=R_{22}-R_{11}=\langle\rho_{22}
\rangle_{\rm sp}-\langle\rho_{11}\rangle_{\rm sp}$,
where $\langle \,\cdots\, \rangle_{\rm sp} $ corresponds to a spatial
average over a volume of the order of a resonance wavelength cubed and
the $\rho_{ij}$ are the density matrix elements for a two-level system
with a lower state $|1\rangle$ and an upper state $|2\rangle$.
Also, $\mu$ is the matrix element of the transition dipole moment,
assumed real, $\gamma_{\perp}$ is a phenomenological dipole dephasing
rate, $\gamma_{\parallel}$ is a phenomenological population relaxation
rate, and $W_{\rm eq}$ is the population difference at equilibrium.
\par
For a linearly polarizable material, the polarization is
${\bf P}=pN_d{\bf E}_L$.
For atoms embedded in a linearly polarizable material, the polarization
is the sum of the linear and nonlinear components.
Substituting the local field (\ref{EQq2.01}) into the linear component,
we have the polarization envelope
$$
{\cal P}=pN_d\left ( {\cal E}+\frac{4\pi}{3}{\cal P}\right ) +2N\mu R_{21}
$$
where $N$ is the number density of atoms.
Collecting terms in ${\cal P}$ and using the Clausius--Mossotti--Lorentz--Lorenz
relation (\ref{EQq2.02}) yields
\begin{equation}
{\cal P}={\frac{\varepsilon-1}{4\pi}}{\cal E}
+{\frac{\varepsilon+2}{3}} 2N\mu R_{21}.
\label{EQq2.05}
\end{equation}
Substituting the polarization envelope (\ref{EQq2.05}) into the generalized
Bloch equations (\ref{EQq2.04}) produces
\begin{subequations}
\label{EQq2.06}
\begin{equation}
{\frac{\partial R_{21}}{\partial t}} =
i\left (\omega-\omega_0-\frac{4\pi}{3\hbar}
N\mu^2\ell W \right )R_{21}
 -  {\frac{i\mu }{2\hbar}}\ell{\cal E} W
 -  \gamma_{\perp} R_{21},
\label{EQq2.06a}
\end{equation}
\begin{equation}
{\frac{\partial W}{\partial t}} =
- {\frac{i}{\hbar}} \left [
 \mu\ell{\cal E}^* R_{21}
-\mu\ell{\cal E} R_{21}^*
 \right ]
-\gamma_{\parallel} (W-W_{\rm eq}) .
\label{EQq2.06b}
\end{equation}
\end{subequations}
For now, we assume that $\ell$ is real.
Then, the local-field effect of the dielectric is simply an enhancement
of the driving field ${\cal E}$, or Rabi frequency $\mu{\cal E}/\hbar$,
and the inversion-dependent Lorentz redshift by $\ell$ \cite{BICBPRA96}.
The decay rates remain phenomenological.
\par
The optical Bloch equations \cite{BIBoyd} are the basic semiclassical
equations of motion for an isolated two-level system in the vacuum.
The two-level system interacts with its environment through the local
field.
Using the classical Lorentz local-field condition shows the effects of
the environment to be:
(i) a Bowden--Lorentz redshift of the resonance frequency by
$(4\pi/3\hbar)N\mu^2 W$ due to nearby atoms,
(ii) a Bloembergen enhancement of the field ${\cal E}$ by $\ell$ due
to the linear host,
and (iii) a Bloembergen-type enhancement of the Bowden--Lorentz
redshift by $\ell$ due to the linear host.
Further consideration of the generalized Bloch equations, derived using
the classical Lorentz local-field condition, is suspended until quantum
electrodynamic equations of motion have been derived.
At that time Eq.\ (\ref{EQq2.06}) will serve as the basis for
quantum--classical correspondence.
\par
\section{Macroscopic Quantum Electrodynamics}
\par
Quantum electrodynamics can be viewed as the quantized version of
Lorentzian electrodynamics in which discrete quantum particles interact
with the vacuum field modes.
When applied to dielectrics, the practice has been to create a
macroscopic version of quantum electrodynamics along the lines of a
quantized version of continuum electrodynamics.
The macroscopic theory can be derived either by quantizing the classical
Maxwell fields or by applying a continuum approximation to the
microscopic quantum electrodynamic Hamiltonian.
Ginzburg \cite{BIGinz} pioneered the procedure of canonical quantization
of the field in a dielectric and applied it to Cherenkov radiation.
The macroscopic quantization procedure was limited to dielectrics with
negligible dispersion and absorption.
Jauch and Watson \cite{BIJauchW,BIJauchWII} continued Ginzburg's work
and extended macroscopic quantization to dispersive
dielectrics \cite{BIMilJMO,BIJauchWIII}, while other researchers have
treated absorption \cite{BIabs2,BIabs3} and nonlinear
dielectrics \cite{BIDrum,BIHilMlo}.
Knoester and Mukamel \cite{BIKnoesM} and Huttner, Baumberg, and Barnett
\cite{BIHBB} start with the fundamental microscopic Hamiltonian and
transform from coordinate space to wave-number space in the continuum
approximation to derive the macroscopic Hopfield
Hamiltonian \cite{BICMomSp}.
Recent work \cite{BIMDLHM0,BIMDLHM1,BIMDLHM2,BIMDLHM3,BIMDLHM4} includes
macroscopic quantization of fields in magnetodielectric media, including
left-handed negative-index materials.
\par
The spontaneous emission rate of an atom in a dielectric is readily
derived by the macroscopic quantum electrodynamic theory.
In 1976, Nienhuis and Alkemade \cite{BINien} used a macroscopic
version of Fermi's golden rule to derive the dielectrically enhanced
spontaneous emission rate with the macroscopic Ginzburg fields.
Huttner, Barnett, and Loudon \cite{BIBHL}, among others, have combined
the macroscopic Fano--Hopfield \cite{BIFano,BIHop} theory
with Fermi's golden rule to derive the dielectric renormalization of the
spontaneous emission rate of an impurity atom, while other studies begin
with macroscopic Green's functions \cite{BIWelsch} or auxiliary
fields \cite{BITip}.
For the most part, the collective attention is focused on the dielectric
renormalization of the spontaneous emission rate.
Knoester and Mukamel \cite{BIKnoesM} obtained the dielectric effect on
both the Lorentz redshift and the spontaneous decay rate by deriving
operator equations of motion from the Hopfield model.
In this section, we derive equations of motion using Weisskopf--Wigner
theory applied to the macroscopic Hamiltonian in terms of the Ginzburg
field operators \cite{BIPhysLett}.
A term-by-term comparison with the generalized Bloch equations
that were derived in the preceding section under the Lorentz
local-field condition exposes an extraordinary degree of disagreement
between two known and accepted treatments of the effect of a dielectric
host on the electrodynamics of two-level atoms.
\par
The principal product of the macroscopic quantization theory is the
medium-assisted field operator 
\begin{equation}
\bar {\bf E}=\frac{i\hbar}{n} \sum_{l\lambda}
\sqrt{\frac{2\pi\omega_l}{\hbar V}}
\left ( \bar a_l e^{i{\bf k}_l\cdot{\bf r}} - H.c. \right )
\hat {\bf e}_{{\bf k}_l \lambda},
\label{EQq3.01}
\end{equation}
where $\bar a_l^{\dagger}$ and $\bar a_l$ are the macroscopic creation
and destruction operators for the field modes and $\omega_{l}$ is the
frequency of the field in the mode $l$.
Also, $V$ is the quantization volume, $\hat {\bf e}_{{\bf k}_l\lambda}$
is a unit vector in the direction of the polarization,
and $\lambda$ denotes the state of polarization.
The spontaneous emission rate of an impurity atom in a dielectric can
then be obtained by applying Fermi's golden rule,
\begin{equation}
\Gamma=\frac{2\pi}{\hbar}|\langle f|H_{int}|i\rangle|^2 D,
\label{EQq3.02}
\end{equation}
to the effective interaction Hamiltonian
$H_{\rm int}=-\mu_a \cdot \bar {\bf E}$,
where $i$ labels the initial state and $f$ denotes all available final
states.
As is typically calculated, the dielectric renormalization of the
vacuum spontaneous emission rate of an atom is found to be
$n$ \cite{BIGlaub,BIMilJMO,BINien} due to the
dielectric renormalization of the
electric field operator (\ref{EQq3.01}) by $1/n$,
which is squared, and the $D=n^3$ density-of-states factor.
\par
Local-field effects of a dielectric are suppressed in the macroscopic
quantization procedure and such local-field effects must be introduced 
phenomenologically.
The paradigm that emerged from the propagation studies of Hopf and
Scully \cite{BIHS} and Bloembergen's work \cite{BIBloem} in nonlinear
optics is that the effect of a dielectric host is to multiply each
occurrence of the dipole moment of a two-level atom by a local-field
enhancement factor \cite{BIKnoesM,BIMarcuse,BIBoyd}.
Using the Lorentz virtual cavity model of the local field,
the spontaneous emission rate
\begin{equation}
\Gamma=n\ell^2\frac{4\omega_b^3|\mu_b|^2}{3c^3\hbar}
=n\left (\frac{n^2+2}{3}\right )^2\Gamma_0
\label{EQq3.03}
\end{equation}
for atoms in a dielectric scales as $n^5$ for large $n$.
For an atom in a real cavity, the local-field enhancement factor
is based on the Onsager model, and the modified spontaneous
emission rate
\begin{equation}
\Gamma=n\left (\frac{3n^2}{2n^2+1}\right )^2\Gamma_0
\label{EQq3.04}
\end{equation}
scales as $n$.
A study of local-field effects by de Vries and Lagendijk \cite{BIdeVL}
found that the Lorentz virtual-cavity model is appropriate if the atom
goes into a crystal substitutionally but that the Onsager real-cavity
model should be used for interstitial impurities.
\par
The macroscopic quantum electrodynamic theory can be used to derive
additional consequences of the dielectric host for two-level atoms.
Taking the field in a coherent state, the effective Hamiltonian is
$$
H_{\rm eff}=
\sum_{js} {\frac{\hbar \omega_a}{2}}\sigma_3^{j}
+\hbar\sum_{l\lambda}\omega_{l}\bar a_{l}^{\dagger}\bar a_{l}
$$
$$
-\frac{i\hbar}{n}\sum_{js} \sum_{l\lambda}
\left (g_{l}^{j}\sigma_+^{j}\bar a_{l}e^{i{\bf k}_{l}
\cdot {\bf r}_j} - {g_{l}^{j}}^*  \bar a_{l}^{\dagger}\sigma_-^{j}
e^{-i{\bf k}_{l} \cdot {\bf r}_j} \right )
$$
\begin{equation}
-\frac{i\mu_a}{2}\sum_{j}
\left (
\sigma_+^j\bar{\cal E}e^{-i(\omega_p t-{\bf k}_p\cdot{\bf r}_j)}
-\bar{\cal E}^*\sigma_-^j e^{i(\omega_p t-{\bf k}_p\cdot{\bf r}_j)}
\right ) .
\label{EQq3.05}
\end{equation}
For species $a$,
$\sigma_3^{j}$ is the inversion operator and $\sigma_{\pm}^{j}$ are
the raising and lowering operators for the $j^{th}$ atom,
$g_{l}^{j}=(2\pi\omega_l/\hbar V)^{1/2}\mu_a(\hat{\bf x}_{j}\cdot
\hat {\bf e}_{{\bf k}_{l}\lambda})$
is the coupling between the atom at position
${\bf r}_j$ and the radiation field,
$\hat{\bf x}_j$ is a unit vector in the direction of the dipole moment
at ${\bf r}_j$,
$\omega_a$ is the transition frequency, and
$\mu_a$ is the matrix element of the transition dipole moment.
\par
Except for coefficients of $1/n$, and macroscopic
field-mode operators, the effective Hamiltonian is the same as the
microscopic Hamiltonian for identical two-level atoms in the vacuum.
Equations of motion can then be derived in the same manner, which is the
primary reason for adopting the macroscopic formalism.
The formal integral of the Heisenberg equation of motion for the
field-mode operators is used to eliminate these operators from the
remaining Heisenberg equations of motion.
One obtains
$$
{\frac{d \sigma_-^{j}}{dt}}= -i\omega_a \sigma_-^j(t)
+\frac{\mu_a}{2 \hbar}\sigma_3^j(t)
\bar{\cal E}(t)e^{-i(\omega_p t-{\bf k}_p\cdot{\bf r}_j)}
$$
$$
+\frac{1}{n}\sum_{l\lambda} {g_l^{j}} \sigma_3^{j}(t) \bar a_l(0)
e^{-i(\omega_l t -{\bf k}_l\cdot{\bf r}_j)}
$$
$$
+\frac{1}{n^2}\sum_{l\lambda }
g_l^{j} \sigma_3^{j}(t)
\int_0^t dt^{\prime}
\hskip -0.9pt
e^{-i\omega_l(t-t^{\prime})}
\hskip -0.9pt
\sum_{i\ne j,s}{g_l^{i}}^*
\sigma_-^{i}(t^{\prime})
e^{i{\bf k}_l\cdot({\bf r}_j-{\bf r}_i)}
$$
\begin{equation}
+\frac{1}{n^2}\sum_{l\lambda }
g_l^{j} \sigma_3^{j}(t)
\int_0^t dt^{\prime} e^{-i\omega_l(t-t^{\prime})}
{g_l^{j}}^*
\sigma_-^{j}(t^{\prime})
\label{EQq3.06}
\end{equation}
and a similar equation of motion for the inversion operator.
The procedure to evaluate these terms for two-level atoms in the
vacuum is generally known and will be considered in detail in the
next section.
For now it is sufficient to note that the only differences from the
vacuum case are the coefficients of powers of $n$ and the
$n^3$ renormalization of the density-of-states in a dielectric
in which the sum over modes is evaluated as
\begin{equation}
\sum_{l}\big \{\big \} \rightarrow
n^3\left ( \frac{L}{2\pi c}\right )^3 \int_0^{\infty}
d\omega_l \omega_l^2 \int d\Omega \big \{\big \}
\label{EQq3.07}
\end{equation}
in the mode continuum limit.
Transforming to a rotating frame of reference and performing a 
local spatial average, one obtains the Bloch-like equations
of motion
\begin{subequations}
\label{EQq3.08}
$$
{\frac{\partial R_{21}}{\partial t}}=
i\left (\omega_p-\omega_a-\frac{4\pi}{3\hbar}N\mu_a^2 n W\right )R_{21}
- {\frac{i\mu_a}{2\hbar}} \bar{\cal E} W
$$
\begin{equation}
-n\frac{\Gamma_0}{2} R_{21},
\label{EQq3.08a}
\end{equation}
\begin{equation}
{\frac{\partial W}{\partial t}}=
- \frac{i}{\hbar}\left [
 \mu_a\bar{\cal E}^* R_{21}
-\mu_a\bar{\cal E} R_{21}^*
 \right ]
-n\Gamma_0 (W+1) ,
\label{EQq3.08b}
\end{equation}
\end{subequations}
where $W=\langle\sigma_3\rangle_{\rm sp}$
and $R_{21}=\langle -i\sigma_-\rangle_{\rm sp}$.
\par
Equations (\ref{EQq3.08}) are the equations of motion for a two-level
atom in a dielectric host medium that are derived using macroscopic
quantum electrodynamics.
The effects of the host appear as an enhancement of the decay rates and
the Bowden--Lorentz redshift by a factor of the refractive index $n$,
when compared to the vacuum $n=1$ case.
However, the equations of motion derived using the classical Lorentz
local-field condition, Eqs.\ (\ref{EQq2.06}), display an enhancement of
both the Bowden--Lorentz redshift and the field by a factor of $\ell$.
\par
The fact that local-field effects are suppressed in the macroscopic
theory is well known.
The accepted practice \cite{BIKnoesM,BIMarcuse,BIBoyd} is to 
phenomenologically associate a local-field factor of $\ell$ with each
occurrence of the dipole moment $\mu$ based on the propagation
studies of Hopf and Scully \cite{BIHS} and Bloembergen's
work \cite{BIBloem} in nonlinear optics.
In this case, Eqs.\ (\ref{EQq3.08}), become
\begin{subequations}
\label{EQq3.09}
$$
{\frac{\partial R_{21}}{\partial t}}=
i\left (\omega_p-\omega_a-\frac{4\pi}{3\hbar}N\mu_a^2
n\ell^2 W\right )R_{21}
- {\frac{i\mu_a}{2\hbar}}\ell \bar{\cal E} W
$$
\begin{equation}
-n\ell^2\frac{\Gamma_0}{2} R_{21},
\label{EQq3.09a}
\end{equation}
\begin{equation}
{\frac{\partial W}{\partial t}}=
- \frac{i}{\hbar}\left [
 \mu_a\ell\bar{\cal E}^* R_{21}
-\mu_a\ell\bar{\cal E} R_{21}^*
 \right ]
-n\ell^2\Gamma_0 (W+1) .
\label{EQq3.09b}
\end{equation}
\end{subequations}
The ad hoc local-field correction gives the Bloembergen
enhancement of the field that was derived classically in
Eq.\ (\ref{EQq2.06}), although the redshift is over-corrected.
More significantly, the Bowden--Lorentz redshift retains the extraneous
factor of the refractive index due to the macroscopically quantized
fields.
Then the equations of motion for two-level atoms in a dielectric host,
Eqs.\ (\ref{EQq3.08}) and (\ref{EQq3.09}), that
were derived using the macroscopically quantized fields are inconsistent
with the generalized Bloch equations (\ref{EQq2.06}) of the preceding
section, and the application of quantized macroscopic fields to the
electrodynamics of two-level atoms is contraindicated by classical
Lorentz
local-field theory.
\par
\section{Microscopic Quantum Electrodynamics}
\par
In the preceding two sections, we derived generalized Bloch
equations of motion for two-level atoms in a dielectric host by two
well-known methods and obtained contradictory results.
The most fundamental theoretical approach is to represent both the
atoms and the dielectric microscopically
\cite{BIKnoesM,BIHoK,BIHB,BIJuz,BISutt,BIKSWelsch,
BIBHL,BIDrum,BIHilMlodIII,BIHBB}
 and derive
equations of motion from first principles.
We show that the results of microscopic quantum electrodynamics affirm
the Lorentz local-field theory with respect to the field renormalization
and the Lorentz redshift.
The macroscopic quantum electrodynamic theory produces different
results for these effects and is therefore not valid.
\par
A dielectric host containing one or more two-level atoms is modeled
quantum electromagnetically as a mixture of two species of atoms,
$a$ and $b$, embedded in the vacuum.
To emphasize the symmetry of the local-field interaction,
both species of atoms are initially treated as two-level systems.
Species $b$ is later taken in the harmonic oscillator limit that is
associated with a relatively large detuning from resonance.
Then, the total Hamiltonian is comprised of the Hamiltonians for the
free atoms, the free-space quantized radiation field, and the
interaction of the two-level systems with the free-space quantized
electromagnetic field.
The multipolar and minimal-coupling Hamiltonians are related by a 
canonical transformation and either can be used.
However, due to the canonical transformation, the circumstances of the
rotating-wave approximation (RWA) are different for the two
Hamiltonians \cite{BICMomSp}.
In the typical derivation of the dielectric susceptibility from the
minimal-coupling Hamiltonian, the RWA is invoked implicitly by replacing
polariton eigenenergies with photon
energies \cite{BIHop,BIFano,BIKnoesM}.
We take the direct route and use the multipolar Hamiltonian in
the RWA.
Using a plane-wave expansion of the electromagnetic field
\begin{equation}
{\bf E}=i\hbar \sum_{l\lambda}
\sqrt{\frac{2\pi\omega_l}{\hbar V}}
\left ( a_l e^{i{\bf k}_l\cdot{\bf r}_j} - H.c. \right )
\hat {\bf e}_{{\bf k}_l \lambda},
\label{EQq4.01}
\end{equation}
the multipolar RWA Hamiltonian is \cite{BICBPRL,BIBerMil}
$$
H=
\sum_{js} {\frac{\hbar \omega_a}{2}}\sigma_3^{j}
+\sum_{ns} {\frac{\hbar \omega_b}{2}}\varsigma_3^{n}
+\sum_{l\lambda}\hbar \omega_{l}a_{l}^{\dagger}a_{l}
$$
$$
-i\hbar\sum_{js} \sum_{l\lambda}
\left (g_{l}^{j}\sigma_+^{j}a_{l}e^{i{\bf k}_{l}
\cdot {\bf r}_j} - {g_{l}^{j}}^* a_{l}^{\dagger}\sigma_-^{j} 
e^{-i{\bf k}_{l} \cdot {\bf r}_j} \right )
$$
\begin{equation}
-i\hbar\sum_{ns}\sum_{l\lambda}\left (h_{l}^{n}
\varsigma_+^{n} a_{l}e^{i{\bf k}_{l} \cdot {\bf r}_n}
- {h_{l}^{n}}^*a_{l}^{\dagger} \varsigma_-^{n} 
e^{-i{\bf k}_{l} \cdot {\bf r}_n} \right ),
\label{EQq4.02}
\end{equation}
where $a_{l}^{\dagger}$ and $a_{l}$ are the creation and
destruction operators for the field modes and $\omega_{l}$ is
the frequency of the field in the mode ${l}$ with $k_{l}=\omega_l/c$.
For species $a$,
$\sigma_3^{j}$ is the inversion operator and $\sigma_{\pm}^{j}$ are
the raising and lowering operators for the $j^{th}$ atom,
$g_{l}^{j}=(2\pi\omega_l/\hbar V)^{1/2}\mu_a(\hat{\bf x}_{j}\cdot
\hat {\bf e}_{{\bf k}_{l}\lambda})$
is the coupling between the atom at position
${\bf r}_j$ and the radiation field,
$\hat{\bf x}_j$ is a unit vector in the direction of the dipole moment
at ${\bf r}_j$,
$\omega_a$ is the transition frequency, and
$\mu_a$ is the dipole moment.
For species $b$, $\varsigma_3^{n}$, $\varsigma_{\pm}^{n}$,
${h_{l}^n}$, ${\bf r}_n$, $\omega_b$, and $\mu_b$,
perform the same functions.
Also, $V$ is the quantization volume, $\hat {\bf e}_{{\bf k}_l \lambda}$
is the polarization vector, and $\lambda$ denotes the state of
polarization.
The polarization indices on the variables have been suppressed for
clarity.
In the two-level approximation, the transitions $s=\Delta m\in(-1,0,+1)$
are treated separately and the operators need not carry a specific
value for the magnetic sublevel \cite{BIMK}.
\par
Equations of motion for the material and field-mode operators are
developed in a straightforward manner from the Hamiltonian using the
Heisenberg equation.
We have
\begin{subequations}
\label{EQq4.03}
\begin{equation}
{\frac{d a_l}{dt}}=-i\omega_l a_l
+\sum_{js} {g_l^{j}}^*\sigma_-^{j} e^{-i{\bf k}_l\cdot{\bf r}_j}
+\sum_{ns} {h_l^{n}}^*\varsigma_-^{n} e^{-i{\bf k}_l\cdot{\bf r}_n},
\label{EQq4.03a}
\end{equation}
\begin{equation}
{\frac{d \sigma_-^{j}}{dt}}=-i\omega_a\sigma_-^{j}
+ \sum_{l\lambda } g_l^{j}
\sigma_3^{j} a_l e^{i{\bf k}_l\cdot{\bf r}_j},
\label{EQq4.03b}
\end{equation}
\begin{equation}
{\frac{d \sigma_3^{j}}{dt}}=
-2\sum_{l\lambda}
\left (g_l^{j}\sigma_+^{j}
a_le^{i{\bf k}_l\cdot{\bf r}_j}
+{g_l^{j}}^* a_l^{\dagger}
\sigma_-^{j} e^{-i{\bf k}_l\cdot{\bf r}_j}\right ),
\label{EQq4.03c}
\end{equation}
\begin{equation}
{\frac{d \varsigma_-^{n}}{dt}}
=-i\omega_b\varsigma_-^{n}+\sum_{l\lambda}
{h_l^{n}}
\varsigma_3^{n}a_l e^{i{\bf k}_l\cdot{\bf r}_n},
\label{EQq4.03d}
\end{equation}
\begin{equation}
{\frac{d \varsigma_3^{n}}{dt}}=
-2\sum_{l\lambda } \left (h_l^{n}
\varsigma_+^{n}a_l e^{i{\bf k}_l\cdot{\bf r}_n}
+ {h_l^{n}}^* a_l^{\dagger}
\varsigma_-^{n} e^{-i{\bf k}_l\cdot{\bf r}_n}\right ).
\label{EQq4.03e}
\end{equation}
\end{subequations}
Bloch-like operator equations of motion are obtained by substituting
the formal integral of the field-mode operator equation (\ref{EQq4.03a}),
$$
a_l(t)=a_l(0)e^{-i\omega_l t} +\int_0^t dt^{\prime}
e^{-i\omega_l (t-t^{\prime})}
$$
\begin{equation}
\times
\left (
\sum_{js}{g_l^{j}}^*\sigma_-^{j}(t^{\prime})
e^{-i{\bf k}_l\cdot{\bf r}_j}
+\sum_{ns} {h_l^{n}}^*\varsigma_-^{n}(t^{\prime})
e^{-i{\bf k}_l\cdot{\bf r}_n}
\right ),
\label{EQq4.04}
\end{equation}
into the material operator equations of motion
(\ref{EQq4.03b}),
(\ref{EQq4.03c}),
(\ref{EQq4.03d}),
and (\ref{EQq4.03e}) \cite{BIPolder}.
We transform operator variables to different rotating frames of
reference in which
$\tilde\sigma_-^{j}=
\sigma_-^{j}e^{i\omega_a t}$
and
$\breve\varsigma_-^{n}=
\varsigma_-^{n}e^{i\omega_b t}$
are slowly varying quantities.
Performing the indicated substitution into Eq.\ (\ref{EQq4.03b})
produces
$$
{\frac{d \tilde\sigma_-^{j}}{dt}}=
\sum_{l\lambda} {g_l^{j}} \sigma_3^{j}(t)
e^{i{\bf k}_l \cdot{\bf r}_j}
a_l(0) e^{-i(\omega_l-\omega_a) t}
+\sigma_3^{j}(t)\sum_{l\lambda }
g_l^{j}
$$
$$
\times
\int_0^t dt^{\prime} e^{-i(\omega_l-\omega_a)(t-t^{\prime})}
\sum_{is}{g_l^{i}}^*
\tilde\sigma_-^{i}(t^{\prime})
e^{i{\bf k}_l\cdot({\bf r}_j-{\bf r}_i)}
$$
$$
+e^{-i(\omega_b-\omega_a)t}\sum_{l\lambda}
g_l^{j} \sigma_3^{j}(t)
 \int_0^t dt^{\prime}
 e^{-i(\omega_l-\omega_b)(t-t^{\prime})}
$$
\begin{equation}
\times
\sum_{ns} {h_l^{n}}^*
\breve\varsigma_-^{n} (t^{\prime})
e^{i{\bf k}_l \cdot({\bf r}_j-{\bf r}_n)}
\label{EQq4.05}
\end{equation}
in normal ordering with ${\bf r}_{ji}={\bf r}_{j}-{\bf r}_{i}$.
Likewise, one obtains
$$
{\frac{d \breve\varsigma_-^{n}}{dt}}
=\sum_{l\lambda}{h_l^{n}}
\varsigma_3^{n}e^{i{\bf k}_l\cdot{\bf r}_n}a_l(0)
e^{-i(\omega_l-\omega_b)t}
$$
$$
+e^{i(\omega_b-\omega_a) t}\sum_{l\lambda} {h_l^{n}} \varsigma_3^{n}(t)
\int_0^t dt^{\prime} e^{-i(\omega_l-\omega_a)(t-t^{\prime})}
$$
$$
\times\sum_{js}{g_l^{j}}^*
\tilde\sigma_-^{j}(t^{\prime})
e^{i{\bf k}_l\cdot({\bf r}_n-{\bf r}_j)}
$$
$$
+\sum_{l\lambda} {h_l^{n}} \varsigma_3^{n}(t)
\int_0^t dt^{\prime} e^{-i(\omega_l-\omega_b)(t-t^{\prime})}
$$
\begin{equation}
\times \sum_{ms} {h_l^{m}}^*
\breve\varsigma_-^{m} (t^{\prime})
e^{i{\bf k}_l\cdot({\bf r}_n-{\bf r}_m)}
\label{EQq4.06}
\end{equation}
from Eq.\ (\ref{EQq4.03d}).
Equations of motion for the inversion operators are obtained from
Eqs.\ (\ref{EQq4.03c}) and (\ref{EQq4.03e}) in a similar fashion.
\par
The field that drives an atom, Eq.\ (\ref{EQq4.04}), consists of the
vacuum field, the self-field, and the reaction field and we can identify
the terms on the right-hand side of Eqs.\ (\ref{EQq4.05}) and
(\ref{EQq4.06}) with fluctuations due to the vacuum field, spontaneous
decay from the self-field, near-dipole--dipole interactions between
same-species atoms associated with the reaction field, and
near-dipole--dipole interactions of an atom with the atoms of the
other species, also associated with the reaction field.
The usual procedure is to limit consideration to only the spontaneous
decay rate of a single impurity atom by dropping the fluctuations and
the single-species interactions for both the bath and impurity atoms.
These terms are retained here because they contain significant
information about local-field effects in dielectrics.
For concreteness, we take species $b$ to be the bath atoms and species
$a$ to be the two-level impurity atoms.
\par
\section{Near-dipole--dipole interaction}
\par
The near-dipole--dipole interaction is the basic mechanism 
of the action of the local field.
The Weisskopf--Wigner-based procedure to evaluate the dipole--dipole
interaction for a dense collection of identical two-level atoms was
developed by Ben-Aryeh, Bowden and Englund \cite{BIBBE}, with
corrections by Benedict, Malyshev, Trifonov, and Zaitsev \cite{BIBMTZ},
to investigate single-species intrinsic optical bistability.
The results apply to both species of two-level atoms, individually,
but we work with species $b$ in order to maintain consistent notation
when we take the harmonic oscillator limit of a two-level atom and
derive the Lorentz local-field correction in a dielectric.
\par
We consider a dense collection of identical two-level atoms of species
$b$ in which the atoms are evenly distributed in the vacuum with a
number density
$N_b$. 
The same-species interaction
$$
I_1=\sum_{l\lambda} {h_l^{n}} \varsigma_3^{n}(t)
\int_0^t dt^{\prime} e^{-i(\omega_l-\omega_b)(t-t^{\prime})}
$$
\begin{equation}
\times
\sum_{ms} {h_l^{m}}^*
\breve\varsigma_-^{m} (t^{\prime})
e^{i{\bf k}_l \cdot({\bf r}_n-{\bf r}_m)}
\label{EQq5.01}
\end{equation}
can be extracted from Eq.\ (\ref{EQq4.06}).
\par
The self-interaction of the $n^{th}$ atom with its own reaction field
is characterized by the term ${\bf r}_m= {\bf r}_n$ in the interaction
as a consequence of the relation $\varsigma_3^{n}(t)
\breve\varsigma_-^{n}(t) = - \breve\varsigma_-^{n}(t)$ between Pauli
spin operators for the same atom.
Then
\begin{equation}
I_1^{\rm self}=\sum_{l\lambda} {h_l^{n}} \varsigma_3^{n}(t)
\int_0^t dt^{\prime} e^{-i(\omega_l-\omega_b)(t-t^{\prime})} 
{h_l^{n}}^*
\breve\varsigma_-^{n} (t^{\prime}).
\label{EQq5.02}
\end{equation}
Applying the typical Weisskopf--Wigner
procedure \cite{BIMilonni,BILouisell,BISargent} in the mode continuum
limit, one obtains
\begin{equation}
I_1^{\rm self}=
-\frac{2\omega_b^3|\mu_b|^2}{3\hbar c^3}\breve\varsigma_-^{n}(t)
=- \frac{\gamma_b}{2}\breve\varsigma_-^{n}(t),
\label{EQq5.03}
\end{equation}
where
\begin{equation}
\gamma_b= \frac{4\omega_b^3|\mu_b|^2}{3\hbar c^3}
\label{EQq5.04}
\end{equation}
is the spontaneous decay rate.
For an atom of species $b$, initially in the excited state, $\gamma_b$
is the spontaneous emission rate $\Gamma_0$ into the vacuum.
\par
The pairwise interaction of atoms is carried in the remaining $m\ne n$
part of the summation.
In the Milonni--Knight \cite{BIMK} model of the interaction of two
identical two-level atoms, the strength of the interaction depends on
the separation distance and the magnetic sublevel transition.
Then \cite{BIMK}, 
\begin{equation}
F_1(R)=e^{iR} \left (-\frac{i}{R}+\frac{i}{R^3}+\frac{1}{R^2} \right )
\label{EQq5.05}
\end{equation}
for $\Delta m=\pm 1$ transitions and
\begin{equation}
F_2(R)= e^{iR}\left (-\frac{2i}{R^3}-\frac{2}{R^2} \right )
\label{EQq5.06}
\end{equation}
for $\Delta m=0$ transitions, where
\begin{equation}
\beta_b=\frac{2\omega_b^3|\mu_b|^2}{3\hbar c^3},
\label{EQq5.07}
\end{equation}
$R=k_b r_{nm}=\omega_b r_{nm}/c$,
$k_b=|{\bf k}_b|$, ${\bf r}_{nm}={\bf r}_n-{\bf r}_m$,
and $r_{nm}=|{\bf r}_{nm}|$.
Performing the summation over the magnetic sublevels, the pairwise
dipole--dipole interaction can be written as
\begin{equation}
I_1^{\rm dd}=
\varsigma_3^{n}(t)
\frac{3}{2}\beta_b
\sum_{m\ne n}
B_{nm}
\breve\varsigma_-^{m}(t-r_{nm}/c),
\label{EQq5.08}
\end{equation}
where
$$
B_{nm}=
\left [ ({\bf\hat x}_m\cdot{\bf\hat x}_n)-
({\bf\hat x}_m\cdot{\bf\hat n}_{nm})
({\bf\hat x}_n\cdot{\bf\hat n}_{nm})\right ]
F_1(k_br_{nm})
$$
\begin{equation}
+({\bf\hat x}_m\cdot{\bf\hat n}_{nm})
({\bf\hat x}_n\cdot{\bf\hat n}_{nm})F_2(k_br_{nm})
\label{EQq5.09}
\end{equation}
and ${\bf\hat n}_{nm}={\bf r}_{nm}/r_{nm}$ is a unit vector in 
the direction of ${\bf r}_{nm}={\bf r}_n-{\bf r}_m$ \cite{BIBBE,BIBMTZ}.
Further, $B_{nm}$ incorporates a view factor to account for the
arrangement of the dipoles in the volume.
\par
The atoms are evenly distributed with a number density $N_b$.
For the $n^{th}$ atom, the single-species dipole--dipole interaction
is obtained in a summation over all other atoms of species $b$.
In the region near ${\bf r}_n$, the interaction is evaluated by taking 
the location of dipoles as discrete, while the continuum
approximation is applied elsewhere.
Then
$$
I_1^{\rm dd}=
\varsigma_3^n(t) 
\frac{3}{2}\beta_b
\sum_{ m:r_{nm} < \delta } B_{nm} \breve\varsigma_-^m(t-r_{nm}/c)
$$
\begin{equation}
+\varsigma_3^{n}(t)\frac{3}{2}\beta_bN_b\int_{V-V_{\delta}}
B\breve\varsigma_-(t-|{\bf r}|/c)d^3{\bf r},
\label{EQq5.10}
\end{equation}
where $\delta$ is the radius of a small spherical volume
$V_{\delta}$, larger than a cubic wavelength, about the
point ${\bf r}_n$.
For cubic symmetry, the field generated by the localized atoms
${\bf r}_m\ne {\bf r}_n$ in the virtual cavity is zero at the
center \cite{BIJackson,BIBBE,BIBMTZ}.
\par
The atom $n$ is located at the origin of a cylindrical volume of
thickness $L$ and radius $R_0$.
The near-dipole--dipole interaction is obtained by evaluating
the integral
$$
I_1^{\rm dd}\approx \frac{3}{2}\beta_b N_b\varsigma_3^n(t)\int_0^{2\pi}
d\phi \int_{-L/2}^{L/2}dz
\int_{\rho_{min}}^{\rho_{max}}\rho
d\rho{\breve\varsigma}_-
$$
\begin{equation}
\times 
\Bigg \{\left [ 1-\left (\frac{\rho}{r}\right )^2
\cos^2\phi\right ]
F_1 +
\left ( \frac{\rho}{r}\right )^2
\cos^2\phi F_2
\Bigg \},
\label{EQq5.11}
\end{equation}
excluding a volume $(4/3)\pi\delta^3$ about the origin
from the range of integration,
resulting in 
\begin{equation}
I_1^{\rm dd}\approx
\frac{-4\pi i }{k_b^3}
\frac{N_b\omega_b^3|\mu_b|^2}{\hbar c^3}
e^{ik_b \delta}\varsigma_3^{n}(t)
\langle\breve\varsigma_-(t-|{\bf r}|/c)\rangle_{\rm sp}.
\label{EQq5.12}
\end{equation}
In the limit $\delta\rightarrow 0$, the near-dipole--dipole interaction
\begin{equation}
I_1^{\rm dd} =-i \nu_b \varsigma_3^{n} (t) \bar\varsigma_-
\label{EQq5.13}
\end{equation}
remains finite.
Here
\begin{equation}
\nu_b=\frac{4\pi}{3\hbar} N_b|\mu_b|^2
\label{EQq5.14}
\end{equation}
is the strength of the near-dipole--dipole interaction
and
$\bar\varsigma_-=\langle\breve\varsigma_-(t-|{\bf r}|/c)\rangle_{\rm sp}$
represents a spatially averaged quantity.
The details of this calculation can be found in the articles by
Ben-Aryeh, Bowden, and Englund \cite{BIBBE} and
by Benedict, Malyshev, Trifonov and Zaitsev \cite{BIBMTZ}.
\par
The atoms of species $b$ can be treated as harmonic oscillators if
all excitation frequencies are far from resonance with $\omega_b$.
In this limit the atom essentially remains in the ground state such
that $\varsigma_3^n \rightarrow -1$.
Then the near dipole--dipole interaction reduces to the Lorentz
local-field correction, shifting the resonance frequency by
$4\pi N_b|\mu_b|^2/(3\hbar)$.
The microscopic result is in full agreement with the classical Lorentz
local-field correction and has been experimentally validated 
\cite{BIMMSB,BIWGC} by selective reflection of Rb from a sapphire
window.
\par
Finally, all of the results of this section can be applied to the
other species of atom.
Repeating for species $a$ yields
\begin{equation}
I_1^{\rm self}=-\frac{2\omega_a^3|\mu_a|^2}{3\hbar c^3}\tilde\sigma_-^{j}(t)
=- \frac{\gamma_a}{2}\tilde\sigma_-^{j}(t)
\label{EQq5.15}
\end{equation}
\begin{equation}
I_1^{\rm dd}\approx -i \nu_a \sigma_3^{j} (t)
\langle\tilde\sigma_-(t-|{\bf r}|/c)\rangle_{\rm sp}
=-i \nu_a \sigma_3^{j} (t) \bar\sigma_-.
\label{EQq5.16}
\end{equation}
In addition,
\begin{equation}
\gamma_a=2\beta_a= \frac{4\omega_a^3|\mu_a|^2}{3\hbar c^3}
\label{EQq5.17}
\end{equation}
and
\begin{equation}
\nu_a=\frac{4\pi}{3\hbar} N_a|\mu_a|^2 
\label{EQq5.18}
\end{equation}
are defined for later use.
\par
\section{Interspecies interaction}
\par
The effect of the interspecies near-dipole--dipole interaction can
also be evaluated microscopically using Weisskopf--Wigner theory.
A single rotating frame of reference is used for both species of
atoms by making the transformation
${\tilde\varsigma}_-^{n}=\breve\varsigma_-^{n}
e^{-i(\omega_b-\omega_a) t}$.
Applying the results of the preceding section, we have
$$
{\frac{d {\tilde\varsigma}_-^{n}}{dt}}
=-i(\omega_b-\omega_a) {\tilde\varsigma}_-^n-\frac{\mu_b}{\hbar}f^-_b
+i\nu_b{\bar\varsigma}_-
-\frac{\gamma_b}{2}{\tilde\varsigma}_-^n
$$
\begin{equation}
-\sum_{l\lambda} {h_l^{n}} 
\int_0^t dt^{\prime} e^{-i(\omega_l-\omega_a)(t-t^{\prime})}
\sum_{js} {g_l^{j}}^*
\tilde\sigma_-^{j} (t^{\prime})
e^{i{\bf k}_l \cdot({\bf r}_n-{\bf r}_j)}
\label{EQq6.01}
\end{equation}
in the harmonic oscillator limit $\varsigma_3^{n}(t)\rightarrow -1$,
where the fluctuating field
$$
f^-_b  = 
\sum_{l\lambda}(2\pi\omega_l\hbar /V)^{1/2}
(\hat{\bf x}_{n}\cdot{\bf e}_{k})
e^{i{\bf k}_l\cdot{\bf r}_n}a_l(0)
e^{-i(\omega_l-\omega_b)t}
$$
is associated with the spontaneous decay rate by the
Kramers--Kronig relations.
\par
The interspecies interaction in Eq.\ (\ref{EQq6.01}) describes how a
specific host atom $n$ interacts pairwise with each of the impurity
atoms.
The formal integral of the equation of motion of the host atoms,
Eq.\ (\ref{EQq6.01}), is
$$
\tilde\varsigma_-^n(t)=
-\int_0^t dt^{\prime} e^{-i\alpha(t-t^{\prime})}\frac{\mu_b}{\hbar}f^-
-\int_0^t dt^{\prime} e^{-i\alpha(t-t^{\prime})}
$$
\begin{equation}
\times
\sum_{l\lambda} h_l^n \int_0^{t^{\prime}} dt^{\prime\prime}
e^{-i(\omega_l-\omega_a)(t^{\prime}-t^{\prime\prime})}
\sum_{is} {g_l^i}^*
\tilde\sigma_-^{i} (t^{\prime\prime})
e^{i{\bf k}_l \cdot{\bf r}_{ni}},
\label{EQq6.02}
\end{equation}
where $\alpha=\omega_b-\omega_a-\nu_b-i\gamma_b/2$.
Substituting Eq.\ (\ref{EQq6.02}) into Eq.\ (\ref{EQq4.05}), one obtains
\begin{equation}
{\frac{d \tilde\sigma_-^{j}}{dt}}
=\frac{\mu_a}{\hbar}\sigma_3^j f^- -i\nu_a\sigma_3^j\bar\sigma_-
-\frac{\gamma_a}{2}\tilde\sigma_-^j
+I_2,
\label{EQq6.03}
\end{equation}
where
$$
I_2=- \sum_{nsl\lambda} g_l^{j} \sigma_3^{j}(t)
\int_0^t dt^{\prime} e^{-i(\omega_l-\omega_a)(t^{\prime}-t)}
{h_l^{n}}^*
$$
$$
\times
 e^{i{\bf k}_l\cdot{\bf r}_{jn}}
\int_0^{t^{\prime}}dt^{\prime\prime}
e^{-i\alpha(t^{\prime}-t^{\prime\prime})}
\sum_{{l^{\prime}}\lambda^{\prime}} h_{l^{\prime}}^n
\int_0^{t^{\prime\prime}} dt^{\prime\prime\prime}
$$
\begin{equation}
\times
e^{-i(\omega_{l^{\prime}}-\omega_a)
(t^{\prime\prime}-t^{\prime\prime\prime})}
\sum_{is^{\prime}} {g_{l^{\prime}}^i}^*
\tilde\sigma_-^{i} (t^{\prime\prime\prime})
e^{i{\bf k}_{l^{\prime}} \cdot{\bf r}_{ni}}.
\label{EQq6.04}
\end{equation}
The term containing the fluctuations has been dropped from consideration
because the procedures presented here apply only to slowly varying
quantities and because there will be no contribution from the 
random fluctuations after averaging.
\par
The direct dipole--dipole interactions between impurity atoms was
derived in Section V.
Equation (\ref{EQq6.04}) contains two such dipole--dipole interactions
between non-identical atoms that are integrated over the different
subspaces corresponding to (i) bath atoms and (ii) two-level
impurity atoms.
Due to the relation $\sigma_3^j(t)\sigma_-^j(t)=-\sigma_-^j(t)$
between Pauli spin operators, the term $i=j$ is the special case that
is associated with the renormalization of the spontaneous decay rate.
This separates the interaction $I_2=I_2^{\rm ndd}+I_2^{\rm self}$ into
$I_2^{\rm self}$ for the case $i=j$ and $I_2^{\rm ndd}$ for the summation over
the rest of the impurity atoms.
The two parts of the interaction will be considered separately.
\par
\subsection{Dielectric Mediated Dipole--Dipole Interaction}
\par
The $n^{th}$ atom of the dielectric interacts pairwise with every
impurity atom.
The term
$$
{I_2^{\rm dd}}^A=\sum_{{l^{\prime}}\lambda^{\prime}}
h_{l^{\prime}}^n \int_0^{t^{\prime\prime}}
dt^{\prime\prime\prime}
e^{-i(\omega_{l^{\prime}}-\omega_a)
(t^{\prime\prime}-t^{\prime\prime\prime})}
$$
\begin{equation}
\times
\sum_{i\ne j, s}
{g_{l^{\prime}}^i}^*
\tilde\sigma_-^{i} (t^{\prime\prime\prime})
e^{i{\bf k}_{l^{\prime}}\cdot({\bf r}_n-{\bf r}_i)},
\label{EQq6.05}
\end{equation}
extracted from Eq.\ (\ref{EQq6.04}), 
can be evaluated in the same manner as in Section V, except that the 
atoms are of different species.
The summation represents the effect of all the impurity atoms
on a single atom of the host material.
The sum over the $i\ne j$ impurity atoms is performed (i)
in the near region by the discrete summation over the impurity atoms and
(ii) elsewhere by treating the impurity atoms in the continuum
limit.
For nonidentical atoms, the pairwise interaction goes
as \cite{BInonid1,BInonid2}
$$
{I_2^{\rm dd}}^A=
\frac{3}{2}\sqrt{\beta_a\beta_b}
\sum_{i:r_{ni}<\delta}B_{ni} \tilde\sigma_-^i(t^{\prime\prime}-r_{ni}/c)
$$
\begin{equation}
+\frac{3}{2}\sqrt{\beta_a\beta_b} N_a\int_{V-V_{\delta}}
B\tilde\sigma_-(t^{\prime\prime}-|{\bf r}|/c) d^3{\bf r}.
\label{EQq6.06}
\end{equation}
Then, Eq.\ (\ref{EQq6.06}) is evaluated as in Sec.\ V to obtain 
\cite{BIBBE}
$$
{I_2^{\rm dd}}^A=-\frac{4\pi i}{3\hbar} N_a \mu_a^* \mu_b 
\langle\tilde\sigma_-(t^{\prime\prime}-|{\bf r}|/c)\rangle_{\rm sp}
$$
\begin{equation}
=-\frac{4\pi i}{3\hbar} N_a \mu_a^* \mu_b \bar\sigma_-.
\label{EQq6.07}
\end{equation}
The quantity
$\bar\sigma_-
=\langle\tilde\sigma_-(t^{\prime\prime}-|{\bf r}|/c)\rangle_{\rm sp}$
is slowly varying in time and the temporal integral
\begin{equation}
{I_2^{\rm dd}}^B(t^{\prime})
=\int_0^{t^{\prime}}dt^{\prime\prime}
e^{-i\alpha(t^{\prime}-t^{\prime\prime})}
\left ( \frac{-4\pi i}{3\hbar}\right ) N_a \mu_a^*\mu_b\bar\sigma_-
\label{EQq6.08}
\end{equation}
can be performed in the adiabatic-following approximation.
Repeatedly integrating Eq.\ (\ref{EQq6.08}) by parts \cite{BICrisp}, the
series can be truncated at the first term in the expansion if the time
rate of change of $\bar\sigma_- $ is much smaller than
$\alpha \bar\sigma_-$ yielding
\begin{equation}
{I_2^{\rm dd}}^B(t^{\prime})=
\left (\frac{-1}{\alpha}\right )\left (\frac{-4\pi i}{3\hbar}\right )
N_a \mu_a^*\mu_b\bar\sigma_- .
\label{EQq6.09}
\end{equation}
\par
The remaining part of the $I_2^{\rm dd}$ integration is another
interspecies dipole--dipole interaction.
In this case, the summation imparts the effect of all the
atoms of the host dielectric, modified by interspecies interaction
with the impurity atoms, on the $j^{th}$ impurity atom.
Combining terms, 
\begin{equation}
{I_2^{\rm dd}}(t^{\prime})=
\frac{-4\pi i}{3\hbar}
N_b \mu_b^*\mu_a
\frac{1}{\alpha}
\frac{4\pi i}{3\hbar}
N_a \mu_a^*\mu_b\sigma_3^j\bar\sigma_- 
\label{EQq6.10}
\end{equation}
becomes
\begin{equation}
{I_2^{\rm dd}}(t^{\prime})=
-\frac{4\pi i}{3}
\chi_b
\nu_a
\sigma_3^j\bar\sigma_- ,
\label{EQq6.11}
\end{equation}
where $\chi_b$ is the linear susceptibility of species $b$.
Adding the direct near-dipole--dipole interaction from
Eq.\ (\ref{EQq6.03}), we obtain
$$
I_2^{\rm dd}-i\nu_a\sigma_3(t)\bar\sigma_- (t)
=-i\left (1+\frac{4\pi}{3}\chi_b\right )
\nu_a\sigma_3(t)\bar\sigma_- (t)
$$
\begin{equation}
= -i\frac{n^2+2}{3}\nu_a\sigma_3(t)\bar\sigma_- (t).
\label{EQq6.12}
\end{equation}
Comparison of Eq.\ (\ref{EQq6.12}) with the single species
dipole--dipole interaction, Eq.\ (\ref{EQq5.16}), shows that the effect
of the dielectric host is to enhance the interaction by a factor of
$\ell=(n^2+2)/3$.
Taking the local spatial average, $W=\langle\sigma_3\rangle_{\rm sp}$
and $R_{21}=\langle -i\sigma_-\rangle_{\rm sp}$, one finds that the Lorentz
redshift
\begin{equation}
\frac{n^2+2}{3}\frac{4\pi}{3\hbar} N\mu^2
\label{EQq6.13}
\end{equation}
is consistent with the Lorentz local-field calculation,
Eq.\ (\ref{EQq2.06a}), while the macroscopic quantum
electrodynamic result, Eq.\ (\ref{EQq3.09a}), is not.
\par
\subsection{Dielectric-Enhanced Spontaneous Decay Rate}
\par
Most of the elements of the microscopic theory of the spontaneous
decay rate of an atom in a dielectric are common to the treatment
of the dipole-dipole interaction.
In order to show this clearly, we consider an equivalent derivation
of the dielectric mediated dipole--dipole interaction.
Performing the temporal integrations first, the interspecies interaction
(\ref{EQq6.04}) can be written as
$$
I_2=-
\sum_{nsl\lambda}
g_l^{j} \sigma_3^{j}(t)
\pi\delta(\omega_l-\omega_a)
{h_l^{n}}^* 
 e^{i{\bf k}_l\cdot({\bf r}_j-{\bf r}_n)}
$$
\begin{equation}
\times
\left ( \frac{-i}{\alpha}\right ) 
\sum_{is^{\prime}l^{\prime}\lambda^{\prime}}
h_{l^{\prime}}^n
\pi\delta(\omega_{l^{\prime}}-\omega_a)
\sum_{is^{\prime}} {g_{l^{\prime}}^i}^*
\tilde\sigma_-^{i} (t^{\prime})
e^{i{\bf k}_{l^{\prime}} \cdot({\bf r}_n-{\bf r}_i) }.
\label{EQq6.14}
\end{equation}
Applying the Milonni-Knight interaction with a view factor results in
$$
I_2=-\frac{-i}{\alpha} \frac{9}{4}\beta_a\beta_b\sigma_3^{j}
\sum_{n}
\Big \{
[({\bf\hat x}_j\cdot{\bf\hat n}_{nj})
({\bf\hat x}_n\cdot{\bf\hat n}_{nj})]
F_2(k_ar_{nj})
$$
$$
+\left [ ({\bf\hat x}_j\cdot{\bf\hat x}_n)-
({\bf\hat x}_j\cdot{\bf\hat n}_{nj})
({\bf\hat x}_n\cdot{\bf\hat n}_{nj})\right ]
F_1(k_ar_{nj})
\Big \}
$$
$$
\times
\sum_{i}
\Big \{
[({\bf\hat x}_i\cdot{\bf\hat n}_{ni})
({\bf\hat x}_n\cdot{\bf\hat n}_{ni})]
F_2(k_ar_{ni})
$$
\begin{equation}
+\left [ ({\bf\hat x}_i\cdot{\bf\hat x}_n)-
({\bf\hat x}_i\cdot{\bf\hat n}_{ni})
({\bf\hat x}_n\cdot{\bf\hat n}_{ni})\right ]
F_1(k_ar_{ni})
\Big \}
\sigma_-^i.
\label{EQq6.15}
\end{equation}
Converting the sums, excluding $i=j$, to integrals and integrating over
the subspace of two-level atoms and then over the subspace of
oscillators is equivalent to the derivation of the dielectric mediated
dipole--dipole interaction that was presented in the preceding
subsection.
\par
The renormalization of the spontaneous decay rate of a
dielectric-embedded two-level atom is derived from the interspecies
interaction (\ref{EQq6.04}) in the same fashion by taking the target
atom to be the same as the source atom.
The summation over the two-level atoms is evaluated with the use of the
delta-function $\delta_{ij}$, rather than the integration over the
subspace of two-level atoms.
Likewise, the summation over the magnetic sublevels
invokes $\delta_{ss^{\prime}}$.
Then,
$$
I_2^{\rm self}=\frac{-i}{\alpha} \frac{9}{4}\beta_a\beta_b\sigma_-^{j}
\sum_{n}
\Big \{
[({\bf\hat x}_j\cdot{\bf\hat n}_{nj})
({\bf\hat x}_n\cdot{\bf\hat n}_{nj})]^2
F_2^2(k_ar_{nj})
$$
\begin{equation}
+\left [ ({\bf\hat x}_j\cdot{\bf\hat x}_n)-
({\bf\hat x}_j\cdot{\bf\hat n}_{nj})
({\bf\hat x}_n\cdot{\bf\hat n}_{nj})\right ]^2
F_1^2(k_ar_{nj})
\Big \}.
\label{EQq6.16}
\end{equation}
The microscopic treatments of the spontaneous decay rate in a dielectric
\cite{BIBerMil,BICBPRL} are missing elements of Eq.\ (\ref{EQq6.16}) and
can neither affirm nor contradict the macroscopic theory of quantum
electrodynamics.
Because the dielectric renormalization of the spontaneous decay rate
does not have a classical local-field condition-based analog, we do not
consider it further.
\par
\section{Dielectric-Enhanced Field}
\par
The dielectric has an effect on an applied electromagnetic field that
can also be evaluated microscopically.
Taking the field in a coherent state, the partial Hamiltonian is
$$
H_f= -\frac{i\hbar}{2}\sum_{j}\left (
\Omega_a \sigma_+^{j} e^{-i(\omega_p t-{\bf k}_{p} \cdot {\bf r}_j)}
-\Omega_a^* \sigma_-^{j} e^{i(\omega_p t-{\bf k}_{p} \cdot {\bf r}_j)}
\right )
$$
\begin{equation}
-\frac{i\hbar}{2}\sum_{n}\left (
\Omega_b \varsigma_+^{n} e^{-i(\omega_p t-{\bf k}_{p} \cdot {\bf r}_n)}
-\Omega_a^* \varsigma_-^{n}e^{i(\omega_p t-{\bf k}_{p} \cdot {\bf r}_n)}
\right ),
\label{EQq7.01}
\end{equation}
where $\omega_p$ is the nominal frequency of the field and
$\Omega_a=\mu_a {\cal E}/\hbar$ and
$\Omega_b=\mu_b {\cal E}/\hbar$ are Rabi frequencies.
The total Hamiltonian is now comprised of the Hamiltonians
(\ref{EQq7.01}) and (\ref{EQq4.02}).
Developing Heisenberg equations of motion and eliminating the field-mode
operators and the dielectric operators results in the appearance of
$$
I_3=
\frac{\mu_a}{2\hbar}\sigma_3^j(t){\cal E}(t)
- \sum_{nsl\lambda} g_l^{j} \sigma_3^{j}(t)
\int_0^t dt^{\prime} e^{-i(\omega_l-\omega_p)(t-t^{\prime})}
$$
\begin{equation}
\times
{h_l^{n}}^* e^{i({\bf k}_l-{\bf k}_p)\cdot{\bf r}_{jn}}
\int_0^{t^{\prime}}dt^{\prime\prime}
e^{\alpha(t^{\prime}-t^{\prime\prime})}
\frac{\mu_a}{2\hbar}{\cal E}(t^{\prime\prime})
\label{EQq7.02}
\end{equation}
as an addition to Eq.\ (\ref{EQq6.03}).
Equation (\ref{EQq7.02}) contains the same type of interaction that was
evaluated in the previous section.
Performing the adiabatic-following approximation and the sum over
polarizations, bath atoms, and magnetic sublevels in the mode 
continuum limit, we obtain
\begin{equation}
I_3=\frac{n^2+2}{3}\frac{\mu_a}{2\hbar}{\cal E} \sigma_3^j.
\label{EQq7.03}
\end{equation}
The electromagnetic field is enhanced by the same factor of $\ell$ as
the reaction field.
\par
\section{Optical Bloch Equations for Embedded Atoms}
\par
The macroscopic optical Bloch equations can be derived from the quantum
electrodynamic equations of motion in the limit of large numbers.
Combining Eqs.\ (\ref{EQq6.03}), (\ref{EQq6.11}),
and (\ref{EQq7.02}), one obtains 
\begin{equation}
{\frac{d\tilde\sigma_-^{j}}{dt}}
= -i\ell\nu_a\sigma_3^j\bar{\sigma}_-
+\ell\frac{\mu_a}{2\hbar}{\cal E}e^{-i(\omega_p-\omega_a) t} \sigma_3^j,
\label{EQq8.01}
\end{equation}
neglecting the Gaussian noise source with zero mean and absorption.
The equation of motion for the inversion operator
\begin{equation}
\frac{d\sigma_3^j}{dt}=
2\Bigg [ i\ell\nu_a\tilde\sigma_+^j\bar{\sigma}_-
-\frac{\mu_a}{2\hbar}\tilde\sigma_+^j
\ell{\cal E}e^{-i(\omega_p-\omega_a)t}
+H.c.\Bigg ]
\label{EQq8.02}
\end{equation}
is derived in a similar manner.
Optical Bloch equations of motion are obtained by 
transforming to a frame rotating at the frequency of the field and
taking a local-spatial average, as in Sec.\ III.
We compare the optical Bloch equations
\begin{subequations}
\label{EQq8.03}
\begin{equation}
{\frac{\partial R_{21}}{\partial t}}  =
i
\left (\omega_p-\omega_a-\frac{4\pi}{3\hbar}N\mu_a^2\ell W
 \right )
R_{21}
- {\frac{i\mu_a }{2\hbar}}\ell{\cal E} W,
\label{EQq8.03a}
\end{equation}
\begin{equation}
{\frac{\partial W}{\partial t}}=
- {\frac{i}{\hbar}} \left [
 \mu_a\ell^*{\cal E}^* R_{21}
-\mu_a\ell{\cal E} R_{21}^*
 \right ],
\label{EQq8.03b}
\end{equation}
\end{subequations}
that were derived from first principles,
to the Lorentz local-field-based equations (\ref{EQq2.06}).
Based on a favorable comparison of the local-field enhancement of the
Lorentz redshift and the Rabi frequency with the classically derived result,
we can reasonably assert that the microscopic theory, unlike the
macroscopic quantum electrodynamic theory, satisfies the correspondence
principle.
\par
The microscopic theory allows us to consider the more general case of 
of a complex local-field enhancement factor.
Separating the real and imaginary parts of $\ell$, the optical Bloch
equations can be written as
\begin{subequations}
\label{EQq8.04}
$$
{\frac{\partial R_{21}}{\partial t}} =
i\left (
\omega_p-\omega_a
-\frac{4\pi}{3\hbar} N\mu_a^2\ell_r W
\right )R_{21}
$$
\begin{equation}
- {\frac{i\mu_a }{2\hbar}}\ell{\cal E} W
-\frac{4\pi}{3\hbar}
N\mu_a^2\ell_i W  R_{21},
\label{EQq8.04a}
\end{equation}
\begin{equation}
{\frac{\partial W}{\partial t}}=
- {\frac{i}{\hbar}} \left [
 \mu_a\ell^*{\cal E}^* R_{21}
-\mu_a\ell{\cal E} R_{21}^*
 \right ] -4\ell_i\nu_a|R_{21}|^2
\label{EQq8.04b}
\end{equation}
\end{subequations}
with $\ell=\ell_r+i\ell_i$.
The microscopic theory justifies the use of a complex refractive index
in the classical Lorentz local-field condition.
Then Eqs.\ (\ref{EQq8.04}), with phenomenological damping, can be
derived by substituting the polarization (\ref{EQq2.05}) with complex
$n$ into the generalized Bloch equations (\ref{EQq2.04}).
The imaginary part of the Lorentz redshift, derived in this manner, was 
found to be associated with an intrinsic cooperative decay for two-level
atoms in an absorptive host \cite{BICBPRA96}.
This result is confirmed by the microscopic theory.
\par
The optical Bloch equations (\ref{EQq8.04}) for dielectric-embedded
two-level atoms are derived from the microscopic description
of quantum electrodynamics using vacuum-based fields that are known to
satisfy the equal-time commutation relations.
Because the field-mode operators have been eliminated, the equal-time
commutation relations cannot be discussed in the context of the optical
Bloch equations (\ref{EQq8.04}) or Heisenberg equations (\ref{EQq8.01}) 
and (\ref{EQq8.02}).
Instead, the optical Bloch equations, generalized for a dielectric
host, must  demonstrate conservation of probability.
The total population is $W^2 +4|R_{21}|^2$.
Direct substitution from Eqs.\ (\ref{EQq8.04}) shows that the temporal
derivative of this quantity is nil, as required, in the limit that
absorption by the atoms and the host dielectric can be neglected.
\par
\section{Summary}
\par
The interest in the dielectric renormalization of the spontaneous
emission rate of an atom embedded in a dielectric material has obscured
the inconsistencies in the macroscopic theory of quantum
electrodynamics.
The dielectric renormalization of the Lorentz redshift and the Rabi
frequency, but not the spontaneous decay rate, can be derived using
the classical Lorentz local-field condition providing an independent
check on the validity of macroscopic quantum electrodynamics.
The optical Bloch equations for a dense collection of two-level atoms
in a dielectric host medium were derived
using the classical Lorentz local-field condition
$$
{\bf E}_L={\bf E}+\frac{4\pi}{3}{\bf P}
$$
in Sec.\ II, while in Sec.\ III, a different set of optical Bloch
equations were derived using the macroscopic quantum electrodynamic
theory.
Both derivations are short and uncomplicated and are based on
well-established physical theories, yet lead to contradictory
results for the Lorentz redshift and the Rabi frequency.
If we assume the validity of the Lorentz local-field condition, then
the macroscopic procedure is proven to be incorrect.
Conversely, the validity of the macroscopic quantum electrodynamic
theory would imply that the Lorentz local-field condition is incorrect.
One deciding factor is that the Lorentz local-field correction has
been validated experimentally \cite{BIMMSB,BIWGC}, while the
experimental record for the macroscopic quantum theory has been
inconclusive.
We applied the more fundamental microscopic theory of quantum
electrodynamics to the same problem and demonstrated complete agreement
with classical theory.
The differences in the Rabi frequencies can be reconciled with a
phenomenological local-field factor applied in the macroscopic case,
providing the virtual-cavity model is used.
However, no such facile reconciliation can be provided for the Lorentz
redshift.
We conclude that both the correspondence principle and microscopic
quantum electrodynamics are violated by the macroscopic quantum
electrodynamic theory.
\par

\end{document}